\documentclass[%
 aip,
 apl,%
 amsmath,amssymb,
 reprint,%
superscriptaddress
]{revtex4-1}

\usepackage{graphicx}
\usepackage{dcolumn}
\usepackage{bm}
\usepackage{comment}
\usepackage{siunitx}

\newcommand{\RNum}[1]{\uppercase\expandafter{\romannumeral #1\relax}}

\begin{document}


\title{A Machine Learning Approach for Automated Fine-Tuning \\ of Semiconductor Spin Qubits}

\author{Julian D. Teske}
 \affiliation{JARA-FIT Institute for Quantum Information, Forschungszentrum J\"ulich GmbH and RWTH Aachen University, 52074 Aachen, Germany}

\author{Simon Humpohl}%
\affiliation{JARA-FIT Institute for Quantum Information, Forschungszentrum J\"ulich GmbH and RWTH Aachen University, 52074 Aachen, Germany}%

\author{Ren\'e Otten}
\affiliation{JARA-FIT Institute for Quantum Information, Forschungszentrum J\"ulich GmbH and RWTH Aachen University, 52074 Aachen, Germany}%

\author{Patrick Bethke}
\affiliation{JARA-FIT Institute for Quantum Information, Forschungszentrum J\"ulich GmbH and RWTH Aachen University, 52074 Aachen, Germany}%

\author{Pascal Cerfontaine}
\affiliation{JARA-FIT Institute for Quantum Information, Forschungszentrum J\"ulich GmbH and RWTH Aachen University, 52074 Aachen, Germany}%

\author{Jonas Dedden}
\affiliation{JARA-FIT Institute for Quantum Information, Forschungszentrum J\"ulich GmbH and RWTH Aachen University, 52074 Aachen, Germany}%

\author{Arne Ludwig}
\affiliation{Lehrstuhl f\"ur Angewandte Festk\"orperphysik, Ruhr-Universit\"at Bochum, 44780 Bochum, Germany}%

\author{Andreas D. Wieck}
\affiliation{Lehrstuhl f\"ur Angewandte Festk\"orperphysik, Ruhr-Universit\"at Bochum, 44780 Bochum, Germany}%

\author{Hendrik Bluhm}
\email{bluhm@physik.rwth-aachen.de}
\affiliation{JARA-FIT Institute for Quantum Information, Forschungszentrum J\"ulich GmbH and RWTH Aachen University, 52074 Aachen, Germany}%

\date{\today}

\begin{abstract}
While spin qubits based on gate-defined quantum dots have demonstrated very favorable properties for quantum computing, one remaining hurdle is the need to tune each of them into a good operating regime by adjusting the voltages applied to electrostatic gates.
The automation of these tuning procedures is a necessary requirement for the operation of a quantum processor based on gate-defined quantum dots, which is yet to be fully addressed.
We present an algorithm for the automated fine-tuning of quantum dots, and demonstrate its performance on a semiconductor singlet-triplet qubit in GaAs. 
The algorithm employs a Kalman filter based on Bayesian statistics to estimate the gradients of the target parameters as function of gate voltages, thus learning the system response.
The algorithm's design is focused on the reduction of the number of required measurements. We experimentally demonstrate the ability to change the operation regime of the qubit within 3 to 5 iterations, corresponding to 10 to 15 minutes of lab-time. 

\end{abstract}

\keywords{Kalman filter, automated tuning, semiconductor spin qubits, quantum computation, quantum dots, machine learning}
\maketitle



The following article has been submitted to Applied Physics Letters.

The realization of qubits based on semiconductor quantum dots has reached a point where concrete architectures for large scale quantum processors are being considered\cite{Vandersypenhotdensecoherent}. Recent achievements representative for the state of the art include demonstrations of two-quibt gates \cite{Zajac439,Watson2018, Veldhorst2015}, single qubit fidelities meeting the requirements for error correction\cite{Yoneda2017}, and first steps in operating arrays of quantum dots\cite{Mortemousque2018, Volk2019,Mills2018}. Another important ingredient for scaleup is inter-qubit coupling over extended distances\cite{Mills2018,Fujita2017,Flentje2017}, e.g. via electron shuttling or cavities\cite{Samkharadze2018, Mi156, Stockklauser2017}.

A central starting point for the operation of qubits based on gate-defined quantum dots is the so-called \textit{tuning} of the system, i.e., the procedure of identifying the voltages that need to be applied to the electrostatic gates to capture and to tunnel-couple individual electrons. 
Tuning by a human operator is very time consuming and will be impractical for multi-qubit systems with more than a handful of qubits.
Efficient tuning is particularly pertinent for gate-defined quantum dots because of the large number of tunable dot parameters and gate voltages to control them, but is also relevant for other systems.
For qubits based on quantum dots, tuning is a two-step procedure,
which comprises both the formation of quantum dots and their depletion into the few electron regime -- which we refer to as coarse-tuning -- and, subsequently, the adjustment of parameters which define the operation conditions of the qubit such as the tunnel coupling to leads and between dots. This procedure is referred to as fine-tuning of the qubit.
 
The initial coarse-tuning of gate-defined quantum dots relies mainly on the recognition of certain features in charge stability diagrams (CSD), which reflects the dot occupancy as a function of gate voltages, or transport through the dots\cite{Hanson2007, Wiel2003}. The most common practice is that a human operator interprets these measurements and decides based on experience and intuition how to adjust gate voltages. 
Such manual tuning schemes have been extended to arrays of quantum dots\cite{Volk2019, Mills2018} using virtual gates that compensate for capacitive crosstalk between physical gates. Automated coarse tuning using image processing tools like the Gabor filter and template matching to identify quantum dot signatures has been demonstrated\cite{Baart2016}. Machine learning techniques like convolutional neural networks to quantify characteristics of CSDs by their similarity to simulated or measured reference CSDs were also explored\cite{Kalantre2017}.

To guide the subsequent fine-tuning, one can either use transport measurements as well (e.g., supplementary material of Ref. \onlinecite{Foletti2009}), or a set of pulsed-gate experiments that extracts the parameters of interest\cite{Botzem2018,VanDiepen2018}.
Adjusting gate voltages based on this information is complicated by the nonlinear dependence of the tunnel couplings on the gate voltages. Furthermore, gate-defined quantum dots exhibit a strongly coupled response, meaning that the voltage applied to each electrode has a considerable influence on several chemical potentials and tunnel couplings. "Virtual gates", i.e., fixed linear combinations of gate voltages each of which predominantly affects a single qubit parameter, are well suited to decouple the control of chemical potentials. However, their use is less straightforward for the control of several tunnel couplings, which in addition to crosstalk typically exhibit a strongly nonlinear dependence on gate voltages.  
To tune a single coupling parameter at a time, van Diepen \textit{et al.}  presented a gradient descent procedure that incremented the virtual gate associated with the target parameter until the desired value was reached\cite{VanDiepen2018}.



Here, we propose and experimentally demonstrate an algorithm for the fine-tuning of qubits based on gate-defined quantum dots which exploits machine learning for improving the efficiency of the tuning procedure.
Our algorithm combines a gradient-based optimization with an adapted implementation of a Kalman filter. The latter allows efficient tracking of the gradients of the parameters in the multidimensional voltage space. Each measurement of the parameters at a new point in the voltage space is compared to the previous measurement and used to update the gradient. This approach results in a full automation of the simultaneous tuning of several coupled parameters. When applied to two tunnel couplings while keeping two chemical potentials fixed, it requires only 3 to 5 iterations to change a tunnel coupling by a factor 2, corresponding to 10-15 minutes of lab-time for the parameter extraction procedures currently used.  

\begin{figure}[]
    \centering
    \includegraphics[width=\linewidth]{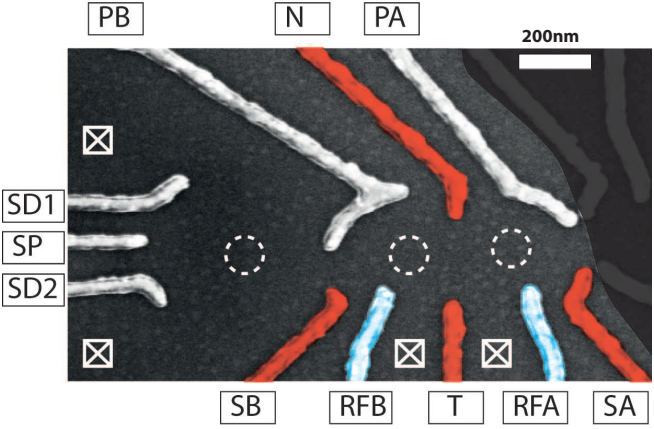}
    \caption{Gate Layout of the sample used for the experimental demonstration. The dashed circles mark the approximate positions of quantum dots. Ohmic contacts are marked by crossed squares. The left dot is used as sensing dot controlled by SD1, SD2 and SP. The RF-gates RFA and RFB, marked in blue, are utilized for rapid control of the chemical potential in the double quantum dot, which is statically set with the plunger gates PA and PB. The gates marked in red are used for tuning the tunnel couplings. Gates T and N are designed to control the inter-dot tunnel coupling whereas SA and SB are meant to control the tunnel coupling to the electron reservoir.}
    \label{fig::gate_layout}
\end{figure}

For the experimental realization  we use a double quantum dot in a AlGaAs/GaAs heterostructure in the same experimental setup as  Botzem \textit{et al.} \cite{Botzem2018} and build on the parameter-extraction procedures developed there. The gate layout is shown in Fig.~\ref{fig::gate_layout}. The double quantum dot is designed to be used as singlet-triplet qubit, and features a neighboring sensing-dot for qubit readout based on spin-to-charge conversion. All measurements are based on RF-reflectometry and performed in a dilution refrigerator.

Although there are 9 DC-gates defining the quantum dots, only 4 of them are used to fine-tune the parameters, namely the gates N, T, SA and SB. The voltages on SD1, SP and SD2 are used to control the sensing dot, and the chemical potentials are controlled with PA and PB. The scans used to extract the parameters of interest are performed with the RF-gates RFA and RFB, which are DC-coupled to an arbitrary waveform generator. The qubit parameters that we want to tune are the strength of the inter-dot tunneling, characterized by the width of the inter-dot transition in gate voltage space, $w$\cite{DiCarlo2004,Botzem2018}, and the time required to reload a singlet $t_{sr}$. The latter characterizes the tunnel coupling between one of the dots and its neighboring electron reservoir and is measured via the dependence of the load efficiency on the corresponding waiting time in the reload operation. Tuning these two parameters is sufficient to obtain a fully operational singlet-triplet qubit \cite{Botzem2018}. The tunnel coupling to the other lead is almost closed to allow for latching readout\cite{Studenikin2012,Mason2015,Harvey-Collard2018}.

\begin{figure}
    \centering
    \includegraphics[width=.8\linewidth]{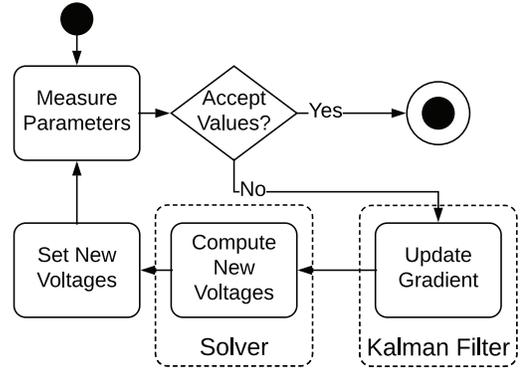}
    \caption{UML activity diagram of the tuning algorithm. The solver can be any gradient based optimization algorithm. The Kalman update consists of Eq.~\eqref{eq::kalman_new_state} and \eqref{eq::kalman_new_cov}. Setting new voltages comprises a compensation of the shift in chemical potential by plunger gates.}
    \label{fig::KalmanSolver}
\end{figure} 

Fine-tuning $w$ and $t_{sr}$ to some target values $w^*$ and $t_{sr}^*$ is complicated by the fact that $w$ and $t_{sr}$ are non-linear functions of four different voltages. To optimize their values, we employ Newton's method with an approximated Jacobian, making our algorithm a quasi-Newton method. In the $k$th iteration of the algorithm, the gate voltages are updated according to the following formula:
\begin{equation}
    \boldsymbol{v}^{(k+1)} = \boldsymbol{v}^{(k)} - \left(\boldsymbol{J}^{(k)}\right)^{-1} \boldsymbol{\Delta}^{(k)}, \label{eq::gaussnewton}
\end{equation}
where $\boldsymbol{v}^{(k)}=(v_N,v_{SA},v_{SB},v_T)^T$ is a vector containing the voltage configuration at step $k$, $\boldsymbol{\Delta}^{(k)}=(t_{sr}^{(k)} - t_{sr}^\ast, w^{(k)} - w^\ast)^T$ contains the distance of the measured parameter values at step $k$ from the target ones, and $\boldsymbol{J}^{(k)}$ is the Jacobian, with $J^{(k)}_{ij} = \partial p_i^{(k)} / \partial v_j^{(k)}$ for $j \in (\text{N},\text{ SA},\text{ SB},\text{ T})$ and $p_i \in (w,t_{\text{sr}})$. $\left(\boldsymbol{J}^{(k)}\right)^{-1} \boldsymbol{\Delta}^{(k)}$ is to be understood as solution of the corresponding, possibly underdetermined system of linear equations with minimal Euclidean norm.

Our tuning procedure combines the Newton step with the Kalman filter as shown in the Unified Modeling Language (UML) diagram in Fig.~\ref{fig::KalmanSolver}. Each iteration begins with a new measurement of the parameters. If the measured values are already in the desired parameter range, the voltages are accepted as final state and the algorithm terminates. Otherwise, gradient updates are performed by a Kalman filter. The new gradients are then used in Eq.~\eqref{eq::gaussnewton} to calculate the new voltages to be applied to the gates. The algorithm mimics the learning process of a human operator, who changes the voltages according to a certain expectation of how this will affect the qubit parameters, performs measurements to verify the result, and uses the information gained from these measurements to refine the understanding of the behavior of the system.

Since Eq.~\eqref{eq::gaussnewton} is based on a linearization of the dependence of $w$ and $t_{sr}$ on the applied voltages, we restrict the maximal voltage change to \SI{10}{\milli \volt}, because the linear approximation by the gradient becomes less accurate the larger the steps are.
On the other hand, the restriction should be chosen as large as possible because the ratio of the physical changes to the fluctuations of the parameters due to noise and disorder grows with the step size. In addition, smaller steps will lead to a larger number of steps for substantial changes. After the new voltages are set, the contrast in the sensing dot is optimized with the gates SD1 and SD2 and the chemical potential is corrected with the plunger gates PA and PB (see supplementary material).


Since the parameters are in general strongly nonlinear functions of the gate voltages, a naive implementation of a gradient-based optimization algorithm would require the Jacobian $\boldsymbol{J}^{(k)}$ to be remeasured by finite differences in every iteration. This is a time-consuming operation because it requires many measurements at different voltages. To avoid this issue, our algorithm takes a machine-learning approach and uses a Kalman filter \cite{Kalman1960,Welch2006}, to estimate the Jacobian at step $k$ using the knowledge of $\boldsymbol{J}^{(k-1)}$ and the information drawn from a single set of measurements. 

The Kalman filter is an algorithm designed to estimate a system of normally distributed random variables given noisy measurements at discrete steps $k$ and knowledge of their dynamics. 
We use an adapted version of the Kalman filter specific to our needs, which are the estimation of the gradient $ g^{(k)}_j = \langle \partial p^{(k)}/\partial v_j \rangle$ and the corresponding covariance matrix $ C_{i,j}^{(k)} = \langle (  \partial p^{(k)}/\partial v_i -  g_i^{(k)}) (  \partial p^{(k)}/\partial v_j -  g_j^{(k)}  ) \rangle$ of a parameter $p$ in iteration $k$ as function of the control voltages $v_{i/j}$ for $i, j \in \left( \text{N},\text{ SA},\text{ SB},\text{ T}\right)$. Expectation values  $\langle \cdot \rangle$ refer to the distribution of the uncertain parameters being tracked. For each $p \in \{ w,t_{\text{sr}} \}$, we use a Kalman filter to track the estimation of its gradient described by $\boldsymbol{g}$ and $\boldsymbol{C}$.
Each instance of the Kalman filter approximates a row in the Jacobian $\boldsymbol{J}$ by the mean of its distribution.
The initial values $\boldsymbol{g}^{(0)}$ and $\boldsymbol{C}^{(0)}$ are measured by finite differences as discussed in the supplementary material.


In each iteration, the Kalman filter uses the information gained from a new measurement of the parameters $p^{(k)}$, to update the values of $\boldsymbol{g}$ and $\boldsymbol{C}$ according to the following formulas\cite{Welch2006}
\begin{align}
    \boldsymbol{g}^{(k)} &= \boldsymbol{g}^{(k-1)} + \boldsymbol{K}^{(k)} (z^{(k)} - \boldsymbol{H}^{(k)}\boldsymbol{g}^{(k-1)}), \label{eq::kalman_new_state}\\
    \boldsymbol{C}^{(k)} &= (I - \boldsymbol{K}^{(k)}\boldsymbol{H}^{(k)})\boldsymbol{C}^{(k-1)} + \boldsymbol{Q}. \label{eq::kalman_new_cov}
\end{align}
Here $\boldsymbol{H}^{(k)}$ is an observation model (consisting in our application of 1 by 4 matrices) that maps the ``state space'' of the Kalman filter (i.e. the space of $\boldsymbol{g}^{(k)}$) onto the measurement space (i.e. the space of $p^{(k)}$), with elements ${H}^{(k)}_{1,j} =  v_j^{(k)} - v_j^{(k-1)}$.  The product $\boldsymbol{H}^{(k)}\boldsymbol{g}^{(k-1)}$ then represents the predicted change in the parameter $p$ due to the change in $\boldsymbol{v}$ at step $k$, which is compared in Eq.~\eqref{eq::kalman_new_state} to the measured change $z^{(k)} = p^{(k)} - p^{(k-1)}$.
The matrix $\boldsymbol{K}^{(k)}$ is the so-called Kalman gain
\begin{equation}
    \boldsymbol{K}^{(k)} = \frac{ \boldsymbol{C}^{(k-1)}\boldsymbol{H}^{(k),T}}{ \boldsymbol{H}^{(k)}\boldsymbol{C}^{(k-1)}\boldsymbol{H}^{(k), T} + {\Delta z^{(k)}}^2 }, \label{eq::kalman_gain}
\end{equation} 
which depends on both the uncertainty of our knowledge of the gradients represented by the covariance matrix $\boldsymbol{C}^{(k)}$, and on the measurement uncertainty ${\Delta z^{(k)}}^2 ={\delta p^{(k)}}^{2} + {\delta p^{(k-1)}}^{2}$, with  $\delta p^{(k)}$ the error on the measurement of $ p^{(k)}$ (see supplementary material).
An inaccurate measurement has a very large ${\Delta z^{(k)}}^2$ and therefore a small gain.  The information gained with the measurement not only contributes to updating the value of the gradient (see Eq.\eqref{eq::kalman_new_state}), but it also reduces the covariance matrix $\boldsymbol{C}^{(k)}$ by a factor determined by the Kalman gain (first term in Eq.~\eqref{eq::kalman_new_cov}). The algorithm becomes Broyden's method in the limit $\Delta z \to 0$. In addition, we include a fixed increased by the term $\boldsymbol{Q}$, which accounts for the additional uncertainty  related to our lack of knowledge how the gradient changes while changing voltages, i.e. of how $\boldsymbol{g}^{(k)}$ deviates from $\boldsymbol{g}^{(k-1)}$.

\begin{figure}[ht!]
    \centering
    \includegraphics[width=\linewidth]{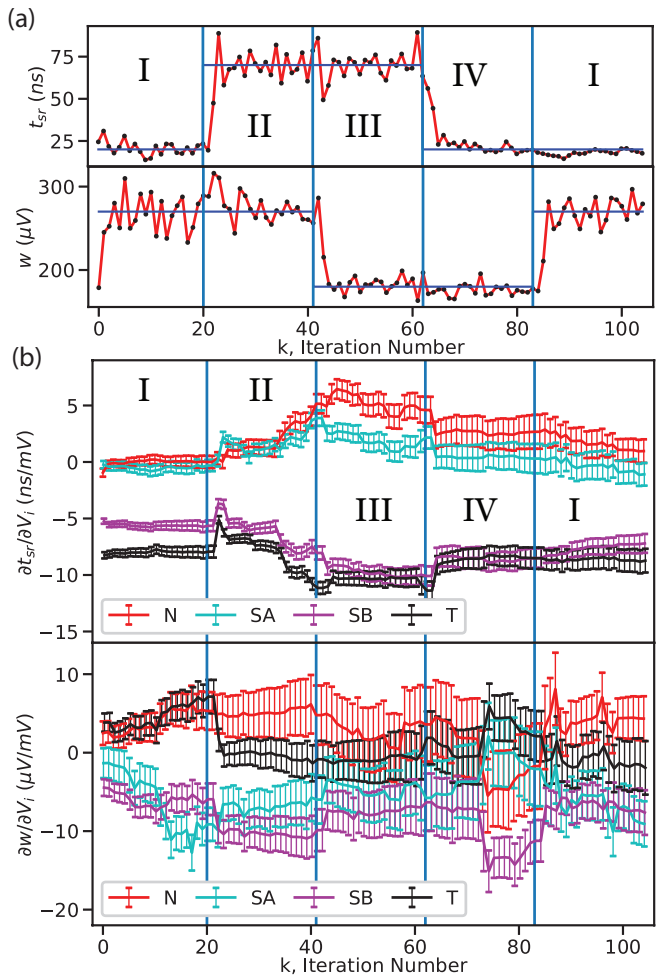}
    \caption{Experimental demonstration of the tuning algorithm. The iteration number $k$ is given by the number of cycles in the diagram shown in Fig.~\ref{fig::KalmanSolver}. (a) The parameters are plotted in red. The horizontal blue lines are the current set points and vertical blue lines mark changes in the set point. Different set points are counted by roman numerals. The parameters are tuned within 3 to 5 iterations into the desired range with an accuracy limited by parameter fluctuations due to statistical noise in the measurement.
    (b) Gradients of the parameters with respect to the voltages on the gates N, SA, SB and T, tracked by the Kalman filter. The error bars are the square roots of the diagonal elements in the gradient's covariance matrix, which are larger in the lower graph because the measurements of the inter-dot transition are generally less accurate. The gradients in the regions with set point \textbf{I} show some differences although the parameters are similar. This can be explained by the influence of the initial gradient estimation and different gate voltages as the system is underdetermined with two parameters controlled by four voltages. (Gate voltages and extension in the supplementary material.)
    }
    \label{fig::benchmarks}
\end{figure}

In our case,  $\boldsymbol{Q}$ is chosen heuristically with the constraint that the estimated increase in the uncertainty of the parameter evaluation, $\boldsymbol{H}^{(k)}\boldsymbol{Q} \boldsymbol{H}^{(k), T}$, must be of the same order of magnitude as the measurement uncertainty ${\Delta z^{(k)}}^2$ for typical values of $\boldsymbol{H}^{(k)}$. This ensures a reasonable Kalman gain $\boldsymbol{K}^{(k)}$ (compare Eq. \eqref{eq::kalman_gain}) and hence change in the prediction $\boldsymbol{g}^{(k)}$. Note that too small values for $\boldsymbol{Q}$ lead to an almost constant $\boldsymbol{g}$, which can cause slow convergence or oscillatory behavior of Eq. \eqref{eq::gaussnewton}. Too large values on the other hand lead to abrupt changes in $\boldsymbol{g}$, which may be problematic if individual measurements occasionally give wrong results, e.g., because the underlying fit does not converge (see supplementary material).

To test the algorithm, we cyclically changed the target parameters $(t_{sr},w)$ in the sequence (\SI{20}{\nano \second}, \SI{270}{\micro  \volt})-(\SI{70}{\nano \second}, \SI{270}{\micro  \volt})-(\SI{70}{\nano \second}, \SI{180}{\micro  \volt})-(\SI{20}{\nano \second}, \SI{180}{\micro  \volt})-(\SI{20}{\nano \second}, \SI{270}{\micro  \volt}).
For each pair of set points, we performed $21$ iterations regardless of the convergence. The resulting voltage changes are not necessarily cyclical as the system is underdetermined.
The choice of the lower set point of $t_{sr}$ is limited by the bandwidth of the data acquisition hardware and the lower set point of $w$ is limited by temperature broadening. We chose the upper limits such that we restrict the range to values typically used in experiments. The observed fluctuations around the setpoints can largely be attributed to noise in the parameter extraction. They are the main limiting factor for the achievable accuracy of the tuning result. According to our experience, this level of accuracy is sufficient for qubit operations, which can tolerate static variations of the parameters considered here by at least 50\%.
The data in Fig.~\ref{fig::benchmarks}(a) demonstrates that the parameters can be tuned individually within three to five steps with a total duration of 10 to 15 minutes. This time includes the compensation of shifts in the chemical potentials and the adjustment of sensor operating points. It is determined by the measurement time, while the computational time is negligible. This performance is comparable to that reported in Ref. \onlinecite{VanDiepen2018} for the tuning of a single tunnel coupling. A speedup of the tuning routine could be achieved by optimizing the way we extract $w$ and $t_{\text{sr}}$. The time required by each measurement is discussed in the supplementary material.



The estimates of the gradients $\boldsymbol{g}^{(k)}$ are plotted in Fig.~\ref{fig::benchmarks}(b).
We see that abrupt changes in the gradient estimates mostly occur in the first steps of each region right after set points are changed. In the upper graph in Fig.~\ref{fig::benchmarks}(b), the absolute values of the elements $\partial t_{sr}/ \partial V_{SB}$ and $\partial t_{sr}/ \partial V_{T}$ are much larger than those of the other two elements because they form the tunnel barrier to the lead next to RFB (see Fig.~\ref{fig::gate_layout}), which is used for the electron exchange in the singlet reload mechanism. The gradient elements $\partial t_{\text{sr}} / \partial v_j$ for  $j \in (N, SA, SB, T)$ are almost constant, indicating only weak effects due to non-linearity.

The gradient of the transition width in the lower graph in Fig.~\ref{fig::benchmarks}(b) is more complicated, presumably because all gates control the positions of the quantum dots and hence the inter-dot tunnel coupling and the transition width. The element $\partial w/ \partial V_{SB}$ changes by a factor of 3 and the other elements change their sign during the experimental demonstration. These changes in the gradients confirm that the dependence of the qubit parameters on the control voltages is strongly non-linear, which indicates that tuning procedures based on pre-calibrated gradients would be inefficient and underpins the advantage of tracking the changes of the gradients during the tuning procedure.



To facilitate the adoption of this approach, we provide an implementation of the algorithm as a python package named qtune\cite{qtune}.
It contains our implementation of the Kalman filter and more tools that simplify an automated fine-tuning program. The implementation complies with good software engineering practices by including a full documentation and unit tests with high coverage. A general interface makes the package adaptable to other setups.

In conclusion, we used the Kalman filter to construct a fully automated fine-tuning procedure.
Thereby, we demonstrated the ability of the Kalman filter to be used in combination with a gradient based optimization algorithm to efficiently solve a non-linear optimization problem without re-measuring the gradient. 
Thus, the algorithm does not only save time and resources, but also provides valuable information about the qubit in form of the gradient of its characteristics as a function of voltages, which can be used for  evaluating its tunability.

Improvements of the convergence behavior could likely be achieved by choosing $\boldsymbol{Q}$ depending on the size of the voltage step in each iteration. In comparison with other quasi-Newton methods like Broyden's methods, our procedure offers the advantage of taking statistical errors into account. The resulting performance advantage is yet to be demonstrated by detailed benchmarks.
Additional performance increases may be achievable by feeding the logarithms of tunnel couplings into the algorithm, thus linearizing the expected exponential dependence of the tunnel coupling on gate voltages\cite{Simmons2009}. However, this would likely not capture all non-linearities, with effects of shifting the electron locations being one potential counterexample.

The algorithm has been tested on a semiconductor spin qubit in a AlGaAs/GaAs double quantum dot but could also be applied to other types of qubit, including any qubit based on gate defined quantum dots. Automated tuning procedures will not only be needed for operating quantum processors but can also be very valuable for the systematic characterization and optimization of qubit technology and reproducibility. In fact, the possibility to tune a qubit is a key criterion for its functionality, and is intimately related to the tuning procedure employed. Hence, "smart" algorithms can make any given qubit design more successful.

{We thank F. Haupt for helpful input on this article and Robert P. G. McNeil for the fabrication of the sample. We acknowledge support by the Impulse and Networking Fund of the Helmholtz association, the Helmholtz Nano Facility (HNF) at the Forschungszentrum J\"ulich\cite{Albrecht2017a}, the Deutsche Forschungsgesellschaft under grant BL 1197/2-1 and BL 1197/4-1 and the Excellence Initiative of the German federal and state governments. A. Ludwig and A. D. Wieck gratefully acknowledge support of DFG-TRR160,  BMBF - Q.Link.X  16KIS0867, and the DFH/UFA  CDFA-05-06.}
\bibliographystyle{apsrev4-1}
\bibliography{library,gitlab}

\begin{thebibliography}{29}%
\makeatletter
\providecommand \@ifxundefined [1]{%
 \@ifx{#1\undefined}
}%
\providecommand \@ifnum [1]{%
 \ifnum #1\expandafter \@firstoftwo
 \else \expandafter \@secondoftwo
 \fi
}%
\providecommand \@ifx [1]{%
 \ifx #1\expandafter \@firstoftwo
 \else \expandafter \@secondoftwo
 \fi
}%
\providecommand \natexlab [1]{#1}%
\providecommand \enquote  [1]{``#1''}%
\providecommand \bibnamefont  [1]{#1}%
\providecommand \bibfnamefont [1]{#1}%
\providecommand \citenamefont [1]{#1}%
\providecommand \href@noop [0]{\@secondoftwo}%
\providecommand \href [0]{\begingroup \@sanitize@url \@href}%
\providecommand \@href[1]{\@@startlink{#1}\@@href}%
\providecommand \@@href[1]{\endgroup#1\@@endlink}%
\providecommand \@sanitize@url [0]{\catcode `\\12\catcode `\$12\catcode
  `\&12\catcode `\#12\catcode `\^12\catcode `\_12\catcode `\%12\relax}%
\providecommand \@@startlink[1]{}%
\providecommand \@@endlink[0]{}%
\providecommand \url  [0]{\begingroup\@sanitize@url \@url }%
\providecommand \@url [1]{\endgroup\@href {#1}{\urlprefix }}%
\providecommand \urlprefix  [0]{URL }%
\providecommand \Eprint [0]{\href }%
\providecommand \doibase [0]{http://dx.doi.org/}%
\providecommand \selectlanguage [0]{\@gobble}%
\providecommand \bibinfo  [0]{\@secondoftwo}%
\providecommand \bibfield  [0]{\@secondoftwo}%
\providecommand \translation [1]{[#1]}%
\providecommand \BibitemOpen [0]{}%
\providecommand \bibitemStop [0]{}%
\providecommand \bibitemNoStop [0]{.\EOS\space}%
\providecommand \EOS [0]{\spacefactor3000\relax}%
\providecommand \BibitemShut  [1]{\csname bibitem#1\endcsname}%
\let\auto@bib@innerbib\@empty
\bibitem [{\citenamefont {Vandersypen}\ \emph {et~al.}(2017)\citenamefont
  {Vandersypen}, \citenamefont {Bluhm}, \citenamefont {Clarke}, \citenamefont
  {Dzurak}, \citenamefont {Ishihara}, \citenamefont {Morello}, \citenamefont
  {Reilly}, \citenamefont {Schreiber},\ and\ \citenamefont
  {Veldhorst}}]{Vandersypenhotdensecoherent}%
  \BibitemOpen
  \bibfield  {author} {\bibinfo {author} {\bibfnamefont {L.~M.~K.}\
  \bibnamefont {Vandersypen}}, \bibinfo {author} {\bibfnamefont
  {H.}~\bibnamefont {Bluhm}}, \bibinfo {author} {\bibfnamefont {J.~S.}\
  \bibnamefont {Clarke}}, \bibinfo {author} {\bibfnamefont {A.~S.}\
  \bibnamefont {Dzurak}}, \bibinfo {author} {\bibfnamefont {R.}~\bibnamefont
  {Ishihara}}, \bibinfo {author} {\bibfnamefont {A.}~\bibnamefont {Morello}},
  \bibinfo {author} {\bibfnamefont {D.~J.}\ \bibnamefont {Reilly}}, \bibinfo
  {author} {\bibfnamefont {L.~R.}\ \bibnamefont {Schreiber}}, \ and\ \bibinfo
  {author} {\bibfnamefont {M.}~\bibnamefont {Veldhorst}},\ }\href@noop {}
  {\bibfield  {journal} {\bibinfo  {journal} {npj Quantum Information}\
  }\textbf {\bibinfo {volume} {3}},\ \bibinfo {pages} {34} (\bibinfo {year}
  {2017})}\BibitemShut {NoStop}%
\bibitem [{\citenamefont {Zajac}\ \emph {et~al.}(2018)\citenamefont {Zajac},
  \citenamefont {Sigillito}, \citenamefont {Russ}, \citenamefont {Borjans},
  \citenamefont {Taylor}, \citenamefont {Burkard},\ and\ \citenamefont
  {Petta}}]{Zajac439}%
  \BibitemOpen
  \bibfield  {author} {\bibinfo {author} {\bibfnamefont {D.~M.}\ \bibnamefont
  {Zajac}}, \bibinfo {author} {\bibfnamefont {A.~J.}\ \bibnamefont
  {Sigillito}}, \bibinfo {author} {\bibfnamefont {M.}~\bibnamefont {Russ}},
  \bibinfo {author} {\bibfnamefont {F.}~\bibnamefont {Borjans}}, \bibinfo
  {author} {\bibfnamefont {J.~M.}\ \bibnamefont {Taylor}}, \bibinfo {author}
  {\bibfnamefont {G.}~\bibnamefont {Burkard}}, \ and\ \bibinfo {author}
  {\bibfnamefont {J.~R.}\ \bibnamefont {Petta}},\ }\href@noop {} {\bibfield
  {journal} {\bibinfo  {journal} {Science}\ }\textbf {\bibinfo {volume}
  {359}},\ \bibinfo {pages} {439} (\bibinfo {year} {2018})}\BibitemShut
  {NoStop}%
\bibitem [{\citenamefont {Watson}\ \emph {et~al.}(2018)\citenamefont {Watson},
  \citenamefont {Philips}, \citenamefont {Kawakami}, \citenamefont {Ward},
  \citenamefont {Scarlino}, \citenamefont {Veldhorst}, \citenamefont {Savage},
  \citenamefont {Lagally}, \citenamefont {Friesen}, \citenamefont
  {Coppersmith}, \citenamefont {Eriksson},\ and\ \citenamefont
  {Vandersypen}}]{Watson2018}%
  \BibitemOpen
  \bibfield  {author} {\bibinfo {author} {\bibfnamefont {T.~F.}\ \bibnamefont
  {Watson}}, \bibinfo {author} {\bibfnamefont {S.~G.}\ \bibnamefont {Philips}},
  \bibinfo {author} {\bibfnamefont {E.}~\bibnamefont {Kawakami}}, \bibinfo
  {author} {\bibfnamefont {D.~R.}\ \bibnamefont {Ward}}, \bibinfo {author}
  {\bibfnamefont {P.}~\bibnamefont {Scarlino}}, \bibinfo {author}
  {\bibfnamefont {M.}~\bibnamefont {Veldhorst}}, \bibinfo {author}
  {\bibfnamefont {D.~E.}\ \bibnamefont {Savage}}, \bibinfo {author}
  {\bibfnamefont {M.~G.}\ \bibnamefont {Lagally}}, \bibinfo {author}
  {\bibfnamefont {M.}~\bibnamefont {Friesen}}, \bibinfo {author} {\bibfnamefont
  {S.~N.}\ \bibnamefont {Coppersmith}}, \bibinfo {author} {\bibfnamefont
  {M.~A.}\ \bibnamefont {Eriksson}}, \ and\ \bibinfo {author} {\bibfnamefont
  {L.~M.}\ \bibnamefont {Vandersypen}},\ }\href {\doibase 10.1038/nature25766}
  {\bibfield  {journal} {\bibinfo  {journal} {Nature}\ }\textbf {\bibinfo
  {volume} {555}},\ \bibinfo {pages} {633} (\bibinfo {year}
  {2018})}\BibitemShut {NoStop}%
\bibitem [{\citenamefont {Veldhorst}\ \emph {et~al.}(2015)\citenamefont
  {Veldhorst}, \citenamefont {Yang}, \citenamefont {Hwang}, \citenamefont
  {Huang}, \citenamefont {Dehollain}, \citenamefont {Muhonen}, \citenamefont
  {Simmons}, \citenamefont {Laucht}, \citenamefont {Hudson}, \citenamefont
  {Itoh}, \citenamefont {Morello},\ and\ \citenamefont
  {Dzurak}}]{Veldhorst2015}%
  \BibitemOpen
  \bibfield  {author} {\bibinfo {author} {\bibfnamefont {M.}~\bibnamefont
  {Veldhorst}}, \bibinfo {author} {\bibfnamefont {C.~H.}\ \bibnamefont {Yang}},
  \bibinfo {author} {\bibfnamefont {J.~C.}\ \bibnamefont {Hwang}}, \bibinfo
  {author} {\bibfnamefont {W.}~\bibnamefont {Huang}}, \bibinfo {author}
  {\bibfnamefont {J.~P.}\ \bibnamefont {Dehollain}}, \bibinfo {author}
  {\bibfnamefont {J.~T.}\ \bibnamefont {Muhonen}}, \bibinfo {author}
  {\bibfnamefont {S.}~\bibnamefont {Simmons}}, \bibinfo {author} {\bibfnamefont
  {A.}~\bibnamefont {Laucht}}, \bibinfo {author} {\bibfnamefont {F.~E.}\
  \bibnamefont {Hudson}}, \bibinfo {author} {\bibfnamefont {K.~M.}\
  \bibnamefont {Itoh}}, \bibinfo {author} {\bibfnamefont {A.}~\bibnamefont
  {Morello}}, \ and\ \bibinfo {author} {\bibfnamefont {A.~S.}\ \bibnamefont
  {Dzurak}},\ }\href {\doibase 10.1038/nature15263} {\bibfield  {journal}
  {\bibinfo  {journal} {Nature}\ }\textbf {\bibinfo {volume} {526}},\ \bibinfo
  {pages} {410} (\bibinfo {year} {2015})}\BibitemShut {NoStop}%
\bibitem [{\citenamefont {Yoneda}\ \emph {et~al.}(2018)\citenamefont {Yoneda},
  \citenamefont {Takeda}, \citenamefont {Ostuka}, \citenamefont {Nakajima},
  \citenamefont {Delbecq},\ and\ \citenamefont {Allison}}]{Yoneda2017}%
  \BibitemOpen
  \bibfield  {author} {\bibinfo {author} {\bibfnamefont {J.}~\bibnamefont
  {Yoneda}}, \bibinfo {author} {\bibfnamefont {K.}~\bibnamefont {Takeda}},
  \bibinfo {author} {\bibfnamefont {T.}~\bibnamefont {Ostuka}}, \bibinfo
  {author} {\bibfnamefont {T.}~\bibnamefont {Nakajima}}, \bibinfo {author}
  {\bibfnamefont {M.~R.}\ \bibnamefont {Delbecq}}, \ and\ \bibinfo {author}
  {\bibfnamefont {G.}~\bibnamefont {Allison}},\ }\href@noop {} {\bibfield
  {journal} {\bibinfo  {journal} {Nature Nanotechnology}\ }\textbf {\bibinfo
  {volume} {13}},\ \bibinfo {pages} {102} (\bibinfo {year} {2018})}\BibitemShut
  {NoStop}%
\bibitem [{\citenamefont {Mortemousque}\ \emph {et~al.}(2018)\citenamefont
  {Mortemousque}, \citenamefont {Chanrion}, \citenamefont {Jadot},
  \citenamefont {Flentje}, \citenamefont {Ludwig}, \citenamefont {Wieck},
  \citenamefont {Urdampilleta}, \citenamefont {Bauerle},\ and\ \citenamefont
  {Meunier}}]{Mortemousque2018}%
  \BibitemOpen
  \bibfield  {author} {\bibinfo {author} {\bibfnamefont {P.-A.}\ \bibnamefont
  {Mortemousque}}, \bibinfo {author} {\bibfnamefont {E.}~\bibnamefont
  {Chanrion}}, \bibinfo {author} {\bibfnamefont {B.}~\bibnamefont {Jadot}},
  \bibinfo {author} {\bibfnamefont {H.}~\bibnamefont {Flentje}}, \bibinfo
  {author} {\bibfnamefont {A.}~\bibnamefont {Ludwig}}, \bibinfo {author}
  {\bibfnamefont {A.~D.}\ \bibnamefont {Wieck}}, \bibinfo {author}
  {\bibfnamefont {M.}~\bibnamefont {Urdampilleta}}, \bibinfo {author}
  {\bibfnamefont {C.}~\bibnamefont {Bauerle}}, \ and\ \bibinfo {author}
  {\bibfnamefont {T.}~\bibnamefont {Meunier}},\ }\href
  {http://arxiv.org/abs/1808.06180} {\  (\bibinfo {year} {2018})},\ \Eprint
  {http://arxiv.org/abs/1808.06180} {arXiv:1808.06180} \BibitemShut {NoStop}%
\bibitem [{\citenamefont {{Volk}}\ \emph {et~al.}(2019)\citenamefont {{Volk}},
  \citenamefont {{Zwerver}}, \citenamefont {{Mukhopadhyay}}, \citenamefont
  {{Eendebak}}, \citenamefont {{van Diepen}}, \citenamefont {{Dehollain}},
  \citenamefont {{Hensgens}}, \citenamefont {{Fujita}}, \citenamefont
  {{Reichl}}, \citenamefont {{Wegscheider}},\ and\ \citenamefont
  {{Vandersypen}}}]{Volk2019}%
  \BibitemOpen
  \bibfield  {author} {\bibinfo {author} {\bibfnamefont {C.}~\bibnamefont
  {{Volk}}}, \bibinfo {author} {\bibfnamefont {A.~M.~J.}\ \bibnamefont
  {{Zwerver}}}, \bibinfo {author} {\bibfnamefont {U.}~\bibnamefont
  {{Mukhopadhyay}}}, \bibinfo {author} {\bibfnamefont {P.~T.}\ \bibnamefont
  {{Eendebak}}}, \bibinfo {author} {\bibfnamefont {C.~J.}\ \bibnamefont {{van
  Diepen}}}, \bibinfo {author} {\bibfnamefont {J.~P.}\ \bibnamefont
  {{Dehollain}}}, \bibinfo {author} {\bibfnamefont {T.}~\bibnamefont
  {{Hensgens}}}, \bibinfo {author} {\bibfnamefont {T.}~\bibnamefont
  {{Fujita}}}, \bibinfo {author} {\bibfnamefont {C.}~\bibnamefont {{Reichl}}},
  \bibinfo {author} {\bibfnamefont {W.}~\bibnamefont {{Wegscheider}}}, \ and\
  \bibinfo {author} {\bibfnamefont {L.~M.~K.}\ \bibnamefont {{Vandersypen}}},\
  }\href@noop {} {\  (\bibinfo {year} {2019})},\ \Eprint
  {http://arxiv.org/abs/1901.00426} {arXiv:1901.00426} \BibitemShut {NoStop}%
\bibitem [{\citenamefont {Mills}\ \emph {et~al.}()\citenamefont {Mills},
  \citenamefont {Zajac}, \citenamefont {Gullans}, \citenamefont {Schupp},
  \citenamefont {Hazard},\ and\ \citenamefont {Petta}}]{Mills2018}%
  \BibitemOpen
  \bibfield  {author} {\bibinfo {author} {\bibfnamefont {A.~R.}\ \bibnamefont
  {Mills}}, \bibinfo {author} {\bibfnamefont {D.~M.}\ \bibnamefont {Zajac}},
  \bibinfo {author} {\bibfnamefont {M.~J.}\ \bibnamefont {Gullans}}, \bibinfo
  {author} {\bibfnamefont {F.~J.}\ \bibnamefont {Schupp}}, \bibinfo {author}
  {\bibfnamefont {T.~M.}\ \bibnamefont {Hazard}}, \ and\ \bibinfo {author}
  {\bibfnamefont {J.~R.}\ \bibnamefont {Petta}},\ }\href@noop {} {\ }\Eprint
  {http://arxiv.org/abs/1809.03976} {arXiv:1809.03976} \BibitemShut {NoStop}%
\bibitem [{\citenamefont {Fujita}\ \emph {et~al.}(2017)\citenamefont {Fujita},
  \citenamefont {Baart}, \citenamefont {Reichl}, \citenamefont {Wegscheider},\
  and\ \citenamefont {K.}}]{Fujita2017}%
  \BibitemOpen
  \bibfield  {author} {\bibinfo {author} {\bibfnamefont {T.}~\bibnamefont
  {Fujita}}, \bibinfo {author} {\bibfnamefont {T.~A.}\ \bibnamefont {Baart}},
  \bibinfo {author} {\bibfnamefont {C.}~\bibnamefont {Reichl}}, \bibinfo
  {author} {\bibfnamefont {W.}~\bibnamefont {Wegscheider}}, \ and\ \bibinfo
  {author} {\bibfnamefont {V.~L.~M.}\ \bibnamefont {K.}},\ }\href@noop {}
  {\bibfield  {journal} {\bibinfo  {journal} {npj Quantum Information}\
  }\textbf {\bibinfo {volume} {3}},\ \bibinfo {pages} {2} (\bibinfo {year}
  {2017})}\BibitemShut {NoStop}%
\bibitem [{\citenamefont {Flentje}\ \emph {et~al.}(2017)\citenamefont
  {Flentje}, \citenamefont {Mortemousque}, \citenamefont {Thalineau},
  \citenamefont {A.}, \citenamefont {Wieck}, \citenamefont {B\"auerle},\ and\
  \citenamefont {Meunier}}]{Flentje2017}%
  \BibitemOpen
  \bibfield  {author} {\bibinfo {author} {\bibfnamefont {H.}~\bibnamefont
  {Flentje}}, \bibinfo {author} {\bibfnamefont {P.-A.}\ \bibnamefont
  {Mortemousque}}, \bibinfo {author} {\bibfnamefont {R.}~\bibnamefont
  {Thalineau}}, \bibinfo {author} {\bibfnamefont {L.}~\bibnamefont {A.}},
  \bibinfo {author} {\bibfnamefont {A.~D.}\ \bibnamefont {Wieck}}, \bibinfo
  {author} {\bibfnamefont {C.}~\bibnamefont {B\"auerle}}, \ and\ \bibinfo
  {author} {\bibfnamefont {T.}~\bibnamefont {Meunier}},\ }\href@noop {}
  {\bibfield  {journal} {\bibinfo  {journal} {npj Quantum Information}\
  }\textbf {\bibinfo {volume} {8}},\ \bibinfo {pages} {501} (\bibinfo {year}
  {2017})}\BibitemShut {NoStop}%
\bibitem [{\citenamefont {Samkharadze}\ \emph {et~al.}(2018)\citenamefont
  {Samkharadze}, \citenamefont {Zheng}, \citenamefont {Kalhor}, \citenamefont
  {Brousse}, \citenamefont {Sammak}, \citenamefont {Mendes}, \citenamefont
  {Blais}, \citenamefont {Scappucci},\ and\ \citenamefont
  {Vandersypen}}]{Samkharadze2018}%
  \BibitemOpen
  \bibfield  {author} {\bibinfo {author} {\bibfnamefont {N.}~\bibnamefont
  {Samkharadze}}, \bibinfo {author} {\bibfnamefont {G.}~\bibnamefont {Zheng}},
  \bibinfo {author} {\bibfnamefont {N.}~\bibnamefont {Kalhor}}, \bibinfo
  {author} {\bibfnamefont {D.}~\bibnamefont {Brousse}}, \bibinfo {author}
  {\bibfnamefont {A.}~\bibnamefont {Sammak}}, \bibinfo {author} {\bibfnamefont
  {U.~C.}\ \bibnamefont {Mendes}}, \bibinfo {author} {\bibfnamefont
  {A.}~\bibnamefont {Blais}}, \bibinfo {author} {\bibfnamefont
  {G.}~\bibnamefont {Scappucci}}, \ and\ \bibinfo {author} {\bibfnamefont
  {L.~M.}\ \bibnamefont {Vandersypen}},\ }\href {\doibase
  10.1126/science.aar4054} {\bibfield  {journal} {\bibinfo  {journal}
  {Science}\ }\textbf {\bibinfo {volume} {359}},\ \bibinfo {pages} {1123}
  (\bibinfo {year} {2018})}\BibitemShut {NoStop}%
\bibitem [{\citenamefont {Mi}\ \emph {et~al.}(2017)\citenamefont {Mi},
  \citenamefont {Cady}, \citenamefont {Zajac}, \citenamefont {Deelman},\ and\
  \citenamefont {Petta}}]{Mi156}%
  \BibitemOpen
  \bibfield  {author} {\bibinfo {author} {\bibfnamefont {X.}~\bibnamefont
  {Mi}}, \bibinfo {author} {\bibfnamefont {J.~V.}\ \bibnamefont {Cady}},
  \bibinfo {author} {\bibfnamefont {D.~M.}\ \bibnamefont {Zajac}}, \bibinfo
  {author} {\bibfnamefont {P.~W.}\ \bibnamefont {Deelman}}, \ and\ \bibinfo
  {author} {\bibfnamefont {J.~R.}\ \bibnamefont {Petta}},\ }\href {\doibase
  10.1126/science.aal2469} {\bibfield  {journal} {\bibinfo  {journal}
  {Science}\ }\textbf {\bibinfo {volume} {355}},\ \bibinfo {pages} {156}
  (\bibinfo {year} {2017})}\BibitemShut {NoStop}%
\bibitem [{\citenamefont {Stockklauser}\ \emph {et~al.}(2017)\citenamefont
  {Stockklauser}, \citenamefont {Scarlino}, \citenamefont {Koski},
  \citenamefont {Gasparinetti}, \citenamefont {Andersen}, \citenamefont
  {Reichl}, \citenamefont {Wegscheider}, \citenamefont {Ihn}, \citenamefont
  {Ensslin},\ and\ \citenamefont {Wallraff}}]{Stockklauser2017}%
  \BibitemOpen
  \bibfield  {author} {\bibinfo {author} {\bibfnamefont {A.}~\bibnamefont
  {Stockklauser}}, \bibinfo {author} {\bibfnamefont {P.}~\bibnamefont
  {Scarlino}}, \bibinfo {author} {\bibfnamefont {J.~V.}\ \bibnamefont {Koski}},
  \bibinfo {author} {\bibfnamefont {S.}~\bibnamefont {Gasparinetti}}, \bibinfo
  {author} {\bibfnamefont {C.~K.}\ \bibnamefont {Andersen}}, \bibinfo {author}
  {\bibfnamefont {C.}~\bibnamefont {Reichl}}, \bibinfo {author} {\bibfnamefont
  {W.}~\bibnamefont {Wegscheider}}, \bibinfo {author} {\bibfnamefont
  {T.}~\bibnamefont {Ihn}}, \bibinfo {author} {\bibfnamefont {K.}~\bibnamefont
  {Ensslin}}, \ and\ \bibinfo {author} {\bibfnamefont {A.}~\bibnamefont
  {Wallraff}},\ }\href@noop {} {\bibfield  {journal} {\bibinfo  {journal}
  {Phys. Rev. X}\ }\textbf {\bibinfo {volume} {7}},\ \bibinfo {pages} {011030}
  (\bibinfo {year} {2017})}\BibitemShut {NoStop}%
\bibitem [{\citenamefont {Hanson}\ \emph {et~al.}(2007)\citenamefont {Hanson},
  \citenamefont {Kouwenhoven}, \citenamefont {Petta}, \citenamefont {Tarucha},\
  and\ \citenamefont {Vandersypen}}]{Hanson2007}%
  \BibitemOpen
  \bibfield  {author} {\bibinfo {author} {\bibfnamefont {R.}~\bibnamefont
  {Hanson}}, \bibinfo {author} {\bibfnamefont {L.~P.}\ \bibnamefont
  {Kouwenhoven}}, \bibinfo {author} {\bibfnamefont {J.~R.}\ \bibnamefont
  {Petta}}, \bibinfo {author} {\bibfnamefont {S.}~\bibnamefont {Tarucha}}, \
  and\ \bibinfo {author} {\bibfnamefont {L.~M.~K.}\ \bibnamefont
  {Vandersypen}},\ }\href {\doibase 10.1103/RevModPhys.79.1217} {\bibfield
  {journal} {\bibinfo  {journal} {Rev. Mod. Phys.}\ }\textbf {\bibinfo {volume}
  {79}},\ \bibinfo {pages} {1217} (\bibinfo {year} {2007})}\BibitemShut
  {NoStop}%
\bibitem [{\citenamefont {van~der Wiel}\ \emph {et~al.}(2002)\citenamefont
  {van~der Wiel}, \citenamefont {De~Franceschi}, \citenamefont {Elzerman},
  \citenamefont {Fujisawa}, \citenamefont {Tarucha},\ and\ \citenamefont
  {Kouwenhoven}}]{Wiel2003}%
  \BibitemOpen
  \bibfield  {author} {\bibinfo {author} {\bibfnamefont {W.~G.}\ \bibnamefont
  {van~der Wiel}}, \bibinfo {author} {\bibfnamefont {S.}~\bibnamefont
  {De~Franceschi}}, \bibinfo {author} {\bibfnamefont {J.~M.}\ \bibnamefont
  {Elzerman}}, \bibinfo {author} {\bibfnamefont {T.}~\bibnamefont {Fujisawa}},
  \bibinfo {author} {\bibfnamefont {S.}~\bibnamefont {Tarucha}}, \ and\
  \bibinfo {author} {\bibfnamefont {L.~P.}\ \bibnamefont {Kouwenhoven}},\
  }\href {\doibase 10.1103/RevModPhys.75.1} {\bibfield  {journal} {\bibinfo
  {journal} {Rev. Mod. Phys.}\ }\textbf {\bibinfo {volume} {75}},\ \bibinfo
  {pages} {1} (\bibinfo {year} {2002})}\BibitemShut {NoStop}%
\bibitem [{\citenamefont {Baart}\ \emph {et~al.}(2016)\citenamefont {Baart},
  \citenamefont {Eendebak}, \citenamefont {Reichl}, \citenamefont
  {Wegscheider},\ and\ \citenamefont {Vandersypen}}]{Baart2016}%
  \BibitemOpen
  \bibfield  {author} {\bibinfo {author} {\bibfnamefont {T.~A.}\ \bibnamefont
  {Baart}}, \bibinfo {author} {\bibfnamefont {P.~T.}\ \bibnamefont {Eendebak}},
  \bibinfo {author} {\bibfnamefont {C.}~\bibnamefont {Reichl}}, \bibinfo
  {author} {\bibfnamefont {W.}~\bibnamefont {Wegscheider}}, \ and\ \bibinfo
  {author} {\bibfnamefont {L.~M.~K.}\ \bibnamefont {Vandersypen}},\ }\href
  {\doibase 10.1063/1.4952624} {\bibfield  {journal} {\bibinfo  {journal}
  {Applied Physics Letters}\ }\textbf {\bibinfo {volume} {108}},\ \bibinfo
  {pages} {213104} (\bibinfo {year} {2016})}\BibitemShut {NoStop}%
\bibitem [{\citenamefont {Kalantre}\ \emph {et~al.}(2017)\citenamefont
  {Kalantre}, \citenamefont {Zwolak}, \citenamefont {Ragole}, \citenamefont
  {Wu}, \citenamefont {Zimmerman}, \citenamefont {Stewart},\ and\ \citenamefont
  {Taylor}}]{Kalantre2017}%
  \BibitemOpen
  \bibfield  {author} {\bibinfo {author} {\bibfnamefont {S.~S.}\ \bibnamefont
  {Kalantre}}, \bibinfo {author} {\bibfnamefont {J.~P.}\ \bibnamefont
  {Zwolak}}, \bibinfo {author} {\bibfnamefont {S.}~\bibnamefont {Ragole}},
  \bibinfo {author} {\bibfnamefont {X.}~\bibnamefont {Wu}}, \bibinfo {author}
  {\bibfnamefont {N.~M.}\ \bibnamefont {Zimmerman}}, \bibinfo {author}
  {\bibfnamefont {M.~D.}\ \bibnamefont {Stewart}}, \ and\ \bibinfo {author}
  {\bibfnamefont {J.~M.}\ \bibnamefont {Taylor}},\ }\href
  {http://arxiv.org/abs/1712.04914} {\  (\bibinfo {year} {2017})},\ \Eprint
  {http://arxiv.org/abs/1712.04914} {arXiv:1712.04914} \BibitemShut {NoStop}%
\bibitem [{\citenamefont {Foletti}\ \emph {et~al.}(2009)\citenamefont
  {Foletti}, \citenamefont {Bluhm}, \citenamefont {Mahalu}, \citenamefont
  {Umansky},\ and\ \citenamefont {Yacoby}}]{Foletti2009}%
  \BibitemOpen
  \bibfield  {author} {\bibinfo {author} {\bibfnamefont {S.}~\bibnamefont
  {Foletti}}, \bibinfo {author} {\bibfnamefont {H.}~\bibnamefont {Bluhm}},
  \bibinfo {author} {\bibfnamefont {D.}~\bibnamefont {Mahalu}}, \bibinfo
  {author} {\bibfnamefont {V.}~\bibnamefont {Umansky}}, \ and\ \bibinfo
  {author} {\bibfnamefont {A.}~\bibnamefont {Yacoby}},\ }\href {\doibase
  10.1038/nphys1424} {\bibfield  {journal} {\bibinfo  {journal} {Nature
  Physics}\ }\textbf {\bibinfo {volume} {5}},\ \bibinfo {pages} {903} (\bibinfo
  {year} {2009})},\ \Eprint {http://arxiv.org/abs/arXiv:1009.5343v1}
  {arXiv:1009.5343v1} \BibitemShut {NoStop}%
\bibitem [{\citenamefont {Botzem}\ \emph {et~al.}(2018)\citenamefont {Botzem},
  \citenamefont {Shulman}, \citenamefont {Foletti}, \citenamefont {Harvey},
  \citenamefont {Dial}, \citenamefont {Bethke}, \citenamefont {Cerfontaine},
  \citenamefont {McNeil}, \citenamefont {Mahalu}, \citenamefont {Umansky},
  \citenamefont {Ludwig}, \citenamefont {Wieck}, \citenamefont {Schuh},
  \citenamefont {Bougeard}, \citenamefont {Yacoby},\ and\ \citenamefont
  {Bluhm}}]{Botzem2018}%
  \BibitemOpen
  \bibfield  {author} {\bibinfo {author} {\bibfnamefont {T.}~\bibnamefont
  {Botzem}}, \bibinfo {author} {\bibfnamefont {M.~D.}\ \bibnamefont {Shulman}},
  \bibinfo {author} {\bibfnamefont {S.}~\bibnamefont {Foletti}}, \bibinfo
  {author} {\bibfnamefont {S.~P.}\ \bibnamefont {Harvey}}, \bibinfo {author}
  {\bibfnamefont {O.~E.}\ \bibnamefont {Dial}}, \bibinfo {author}
  {\bibfnamefont {P.}~\bibnamefont {Bethke}}, \bibinfo {author} {\bibfnamefont
  {P.}~\bibnamefont {Cerfontaine}}, \bibinfo {author} {\bibfnamefont
  {R.~P.~G.}\ \bibnamefont {McNeil}}, \bibinfo {author} {\bibfnamefont
  {D.}~\bibnamefont {Mahalu}}, \bibinfo {author} {\bibfnamefont
  {V.}~\bibnamefont {Umansky}}, \bibinfo {author} {\bibfnamefont
  {A.}~\bibnamefont {Ludwig}}, \bibinfo {author} {\bibfnamefont
  {A.}~\bibnamefont {Wieck}}, \bibinfo {author} {\bibfnamefont
  {D.}~\bibnamefont {Schuh}}, \bibinfo {author} {\bibfnamefont
  {D.}~\bibnamefont {Bougeard}}, \bibinfo {author} {\bibfnamefont
  {A.}~\bibnamefont {Yacoby}}, \ and\ \bibinfo {author} {\bibfnamefont
  {H.}~\bibnamefont {Bluhm}},\ }\href {\doibase
  10.1103/PhysRevApplied.10.054026} {\bibfield  {journal} {\bibinfo  {journal}
  {Phys. Rev. Applied}\ }\textbf {\bibinfo {volume} {10}},\ \bibinfo {pages}
  {054026} (\bibinfo {year} {2018})}\BibitemShut {NoStop}%
\bibitem [{\citenamefont {van Diepen}\ \emph {et~al.}(2018)\citenamefont {van
  Diepen}, \citenamefont {Eendebak}, \citenamefont {Buijtendorp}, \citenamefont
  {Mukhopadhyay}, \citenamefont {Fujita}, \citenamefont {Reichl}, \citenamefont
  {Wegscheider},\ and\ \citenamefont {Vandersypen}}]{VanDiepen2018}%
  \BibitemOpen
  \bibfield  {author} {\bibinfo {author} {\bibfnamefont {C.~J.}\ \bibnamefont
  {van Diepen}}, \bibinfo {author} {\bibfnamefont {P.~T.}\ \bibnamefont
  {Eendebak}}, \bibinfo {author} {\bibfnamefont {B.~T.}\ \bibnamefont
  {Buijtendorp}}, \bibinfo {author} {\bibfnamefont {U.}~\bibnamefont
  {Mukhopadhyay}}, \bibinfo {author} {\bibfnamefont {T.}~\bibnamefont
  {Fujita}}, \bibinfo {author} {\bibfnamefont {C.}~\bibnamefont {Reichl}},
  \bibinfo {author} {\bibfnamefont {W.}~\bibnamefont {Wegscheider}}, \ and\
  \bibinfo {author} {\bibfnamefont {L.~M.~K.}\ \bibnamefont {Vandersypen}},\
  }\href {\doibase 10.1063/1.5031034} {\bibfield  {journal} {\bibinfo
  {journal} {Applied Physics Letters}\ }\textbf {\bibinfo {volume} {113}},\
  \bibinfo {pages} {033101} (\bibinfo {year} {2018})}\BibitemShut {NoStop}%
\bibitem [{\citenamefont {DiCarlo}\ \emph {et~al.}(2004)\citenamefont
  {DiCarlo}, \citenamefont {Lynch}, \citenamefont {Johnson}, \citenamefont
  {Childress}, \citenamefont {Crockett}, \citenamefont {Marcus}, \citenamefont
  {Hanson},\ and\ \citenamefont {Gossard}}]{DiCarlo2004}%
  \BibitemOpen
  \bibfield  {author} {\bibinfo {author} {\bibfnamefont {L.}~\bibnamefont
  {DiCarlo}}, \bibinfo {author} {\bibfnamefont {H.~J.}\ \bibnamefont {Lynch}},
  \bibinfo {author} {\bibfnamefont {A.~C.}\ \bibnamefont {Johnson}}, \bibinfo
  {author} {\bibfnamefont {L.~I.}\ \bibnamefont {Childress}}, \bibinfo {author}
  {\bibfnamefont {K.}~\bibnamefont {Crockett}}, \bibinfo {author}
  {\bibfnamefont {C.~M.}\ \bibnamefont {Marcus}}, \bibinfo {author}
  {\bibfnamefont {M.~P.}\ \bibnamefont {Hanson}}, \ and\ \bibinfo {author}
  {\bibfnamefont {A.~C.}\ \bibnamefont {Gossard}},\ }\href {\doibase
  10.1103/PhysRevLett.92.226801} {\bibfield  {journal} {\bibinfo  {journal}
  {Phys. Rev. Lett.}\ }\textbf {\bibinfo {volume} {92}},\ \bibinfo {pages}
  {226801} (\bibinfo {year} {2004})}\BibitemShut {NoStop}%
\bibitem [{\citenamefont {Studenikin}\ \emph {et~al.}(2012)\citenamefont
  {Studenikin}, \citenamefont {Thorgrimson}, \citenamefont {Aers},
  \citenamefont {Kam}, \citenamefont {Zawadzki}, \citenamefont {Wasilewski},
  \citenamefont {Bogan},\ and\ \citenamefont {Sachrajda}}]{Studenikin2012}%
  \BibitemOpen
  \bibfield  {author} {\bibinfo {author} {\bibfnamefont {S.~A.}\ \bibnamefont
  {Studenikin}}, \bibinfo {author} {\bibfnamefont {J.}~\bibnamefont
  {Thorgrimson}}, \bibinfo {author} {\bibfnamefont {G.~C.}\ \bibnamefont
  {Aers}}, \bibinfo {author} {\bibfnamefont {A.}~\bibnamefont {Kam}}, \bibinfo
  {author} {\bibfnamefont {P.}~\bibnamefont {Zawadzki}}, \bibinfo {author}
  {\bibfnamefont {Z.~R.}\ \bibnamefont {Wasilewski}}, \bibinfo {author}
  {\bibfnamefont {A.}~\bibnamefont {Bogan}}, \ and\ \bibinfo {author}
  {\bibfnamefont {A.~S.}\ \bibnamefont {Sachrajda}},\ }\href {\doibase
  10.1063/1.4749281} {\bibfield  {journal} {\bibinfo  {journal} {Applied
  Physics Letters}\ }\textbf {\bibinfo {volume} {101}},\ \bibinfo {pages}
  {233101} (\bibinfo {year} {2012})}\BibitemShut {NoStop}%
\bibitem [{\citenamefont {Mason}\ \emph {et~al.}(2015)\citenamefont {Mason},
  \citenamefont {Studenikin}, \citenamefont {Kam}, \citenamefont {Wasilewski},
  \citenamefont {Sachrajda},\ and\ \citenamefont {Kycia}}]{Mason2015}%
  \BibitemOpen
  \bibfield  {author} {\bibinfo {author} {\bibfnamefont {J.~D.}\ \bibnamefont
  {Mason}}, \bibinfo {author} {\bibfnamefont {S.~A.}\ \bibnamefont
  {Studenikin}}, \bibinfo {author} {\bibfnamefont {A.}~\bibnamefont {Kam}},
  \bibinfo {author} {\bibfnamefont {Z.~R.}\ \bibnamefont {Wasilewski}},
  \bibinfo {author} {\bibfnamefont {A.~S.}\ \bibnamefont {Sachrajda}}, \ and\
  \bibinfo {author} {\bibfnamefont {J.~B.}\ \bibnamefont {Kycia}},\ }\href
  {\doibase 10.1103/PhysRevB.92.125434} {\bibfield  {journal} {\bibinfo
  {journal} {Phys. Rev. B}\ }\textbf {\bibinfo {volume} {92}},\ \bibinfo
  {pages} {125434} (\bibinfo {year} {2015})}\BibitemShut {NoStop}%
\bibitem [{\citenamefont {Harvey-Collard}\ \emph {et~al.}(2018)\citenamefont
  {Harvey-Collard}, \citenamefont {D'Anjou}, \citenamefont {Rudolph},
  \citenamefont {Jacobson}, \citenamefont {Dominguez}, \citenamefont
  {Ten~Eyck}, \citenamefont {Wendt}, \citenamefont {Pluym}, \citenamefont
  {Lilly}, \citenamefont {Coish}, \citenamefont {Pioro-Ladri\`ere},\ and\
  \citenamefont {Carroll}}]{Harvey-Collard2018}%
  \BibitemOpen
  \bibfield  {author} {\bibinfo {author} {\bibfnamefont {P.}~\bibnamefont
  {Harvey-Collard}}, \bibinfo {author} {\bibfnamefont {B.}~\bibnamefont
  {D'Anjou}}, \bibinfo {author} {\bibfnamefont {M.}~\bibnamefont {Rudolph}},
  \bibinfo {author} {\bibfnamefont {N.~T.}\ \bibnamefont {Jacobson}}, \bibinfo
  {author} {\bibfnamefont {J.}~\bibnamefont {Dominguez}}, \bibinfo {author}
  {\bibfnamefont {G.~A.}\ \bibnamefont {Ten~Eyck}}, \bibinfo {author}
  {\bibfnamefont {J.~R.}\ \bibnamefont {Wendt}}, \bibinfo {author}
  {\bibfnamefont {T.}~\bibnamefont {Pluym}}, \bibinfo {author} {\bibfnamefont
  {M.~P.}\ \bibnamefont {Lilly}}, \bibinfo {author} {\bibfnamefont {W.~A.}\
  \bibnamefont {Coish}}, \bibinfo {author} {\bibfnamefont {M.}~\bibnamefont
  {Pioro-Ladri\`ere}}, \ and\ \bibinfo {author} {\bibfnamefont {M.~S.}\
  \bibnamefont {Carroll}},\ }\href {\doibase 10.1103/PhysRevX.8.021046}
  {\bibfield  {journal} {\bibinfo  {journal} {Phys. Rev. X}\ }\textbf {\bibinfo
  {volume} {8}},\ \bibinfo {pages} {021046} (\bibinfo {year}
  {2018})}\BibitemShut {NoStop}%
\bibitem [{\citenamefont {Kalman}(1960)}]{Kalman1960}%
  \BibitemOpen
  \bibfield  {author} {\bibinfo {author} {\bibfnamefont {R.~E.}\ \bibnamefont
  {Kalman}},\ }\href {\doibase 10.1115/1.3662552} {\bibfield  {journal}
  {\bibinfo  {journal} {Journal of Basic Engineering}\ }\textbf {\bibinfo
  {volume} {82}},\ \bibinfo {pages} {35} (\bibinfo {year} {1960})}\BibitemShut
  {NoStop}%
\bibitem [{\citenamefont {Welch}\ and\ \citenamefont
  {Bishop}(1995)}]{Welch2006}%
  \BibitemOpen
  \bibfield  {author} {\bibinfo {author} {\bibfnamefont {G.}~\bibnamefont
  {Welch}}\ and\ \bibinfo {author} {\bibfnamefont {G.}~\bibnamefont {Bishop}},\
  }\href@noop {} {\emph {\bibinfo {title} {An Introduction to the Kalman
  Filter}}},\ \bibinfo {type} {Tech. Rep.}\ (\bibinfo {address} {Chapel Hill,
  NC, USA},\ \bibinfo {year} {1995})\BibitemShut {NoStop}%
\bibitem [{\citenamefont {Teske}\ and\ \citenamefont {Humpohl}()}]{qtune}%
  \BibitemOpen
  \bibfield  {author} {\bibinfo {author} {\bibfnamefont {J.}~\bibnamefont
  {Teske}}\ and\ \bibinfo {author} {\bibfnamefont {S.}~\bibnamefont
  {Humpohl}},\ }\href@noop {} {\enquote {\bibinfo {title} {{qtune} fine-tuning
  package},}\ }\bibinfo {howpublished}
  {https://github.com/qutech/qtune}\BibitemShut {NoStop}%
\bibitem [{\citenamefont {Simmons}\ \emph {et~al.}(2009)\citenamefont
  {Simmons}, \citenamefont {Thalakulam}, \citenamefont {Rosemeyer},
  \citenamefont {{Van Bael}}, \citenamefont {Sackmann}, \citenamefont {Savage},
  \citenamefont {Lagally}, \citenamefont {Joynt}, \citenamefont {Friesen},
  \citenamefont {Coppersmith},\ and\ \citenamefont {Eriksson}}]{Simmons2009}%
  \BibitemOpen
  \bibfield  {author} {\bibinfo {author} {\bibfnamefont {C.~B.}\ \bibnamefont
  {Simmons}}, \bibinfo {author} {\bibfnamefont {M.}~\bibnamefont {Thalakulam}},
  \bibinfo {author} {\bibfnamefont {B.~M.}\ \bibnamefont {Rosemeyer}}, \bibinfo
  {author} {\bibfnamefont {B.~J.}\ \bibnamefont {{Van Bael}}}, \bibinfo
  {author} {\bibfnamefont {E.~K.}\ \bibnamefont {Sackmann}}, \bibinfo {author}
  {\bibfnamefont {D.~E.}\ \bibnamefont {Savage}}, \bibinfo {author}
  {\bibfnamefont {M.~G.}\ \bibnamefont {Lagally}}, \bibinfo {author}
  {\bibfnamefont {R.}~\bibnamefont {Joynt}}, \bibinfo {author} {\bibfnamefont
  {M.}~\bibnamefont {Friesen}}, \bibinfo {author} {\bibfnamefont {S.~N.}\
  \bibnamefont {Coppersmith}}, \ and\ \bibinfo {author} {\bibfnamefont {M.~A.}\
  \bibnamefont {Eriksson}},\ }\href {\doibase 10.1021/nl9014974} {\bibfield
  {journal} {\bibinfo  {journal} {Nano Letters}\ }\textbf {\bibinfo {volume}
  {9}},\ \bibinfo {pages} {3234} (\bibinfo {year} {2009})}\BibitemShut
  {NoStop}%
\bibitem [{\citenamefont {Albrecht}\ \emph {et~al.}(2017)\citenamefont
  {Albrecht}, \citenamefont {Moers},\ and\ \citenamefont
  {Hermanns}}]{Albrecht2017a}%
  \BibitemOpen
  \bibfield  {author} {\bibinfo {author} {\bibfnamefont {W.}~\bibnamefont
  {Albrecht}}, \bibinfo {author} {\bibfnamefont {J.}~\bibnamefont {Moers}}, \
  and\ \bibinfo {author} {\bibfnamefont {B.}~\bibnamefont {Hermanns}},\ }\href
  {\doibase http://dx.doi.org/10.17815/jlsrf-3-158} {\bibfield  {journal}
  {\bibinfo  {journal} {Journal of large-scale research facilities JLSRF}\
  }\textbf {\bibinfo {volume} {3}},\ \bibinfo {pages} {A112} (\bibinfo {year}
  {2017})}\BibitemShut {NoStop}%
\end{thebibliography}%

\clearpage
\newpage

\begin{center}
    \textbf{Supplementary Material}
\end{center}

The supplements contain an extension of the experimental data in section~\ref{sec::full_benchmarks} and additional information on the tuning procedure in section~\ref{sec::sd_cds}. More instructions for the employment of the Kalman filter can be found in section~\ref{sec::init}, \ref{sec::measurement_noise} and \ref{sec::choice_of_q}.

\section{Extension of Experimental Data}
\label{sec::full_benchmarks}
\begin{figure*}
    \centering
    \includegraphics[width=\linewidth]{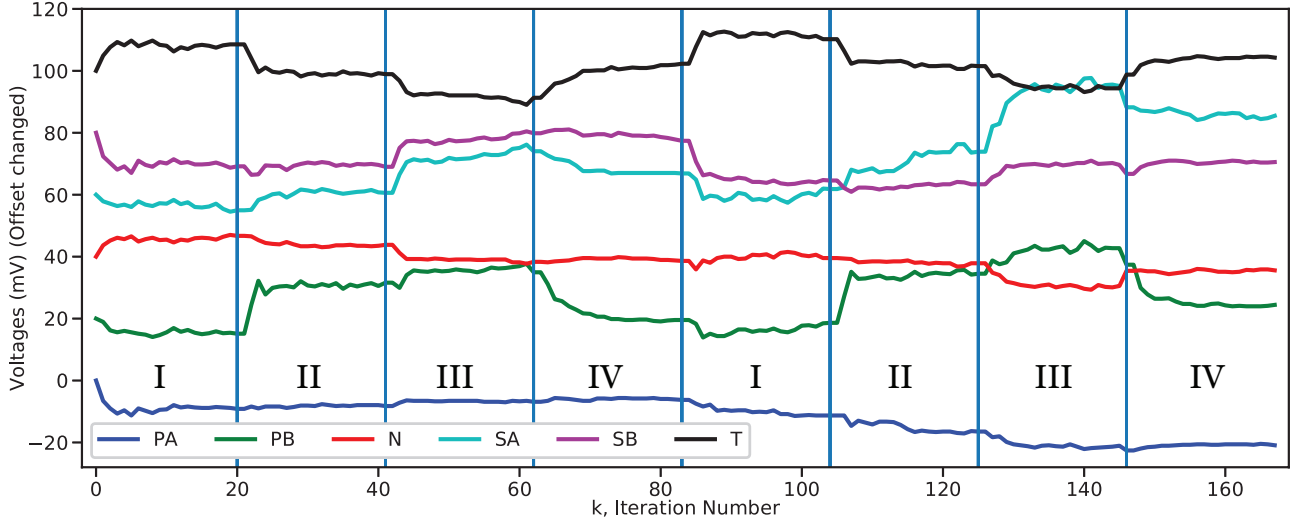}
    \caption{Voltage changes on the dot defining gates. The offsets at the iteration number $k=0$ are chosen as multiples of \SI{20}{\milli \volt} to improve the readability.}
    \label{fig::voltages}
\end{figure*}

We include an extension of the data in Fig.~\ref{fig::second_half} for completeness and to demonstrate the stability of the algorithm. 
The voltages shown in Fig.~\ref{fig::voltages} are not similar for equal set points, which explains different noise levels at similar set points (e.g. for regions \textbf{Ia} and \textbf{Ib}). 



\begin{figure}
    \centering
    \includegraphics[width=\linewidth]{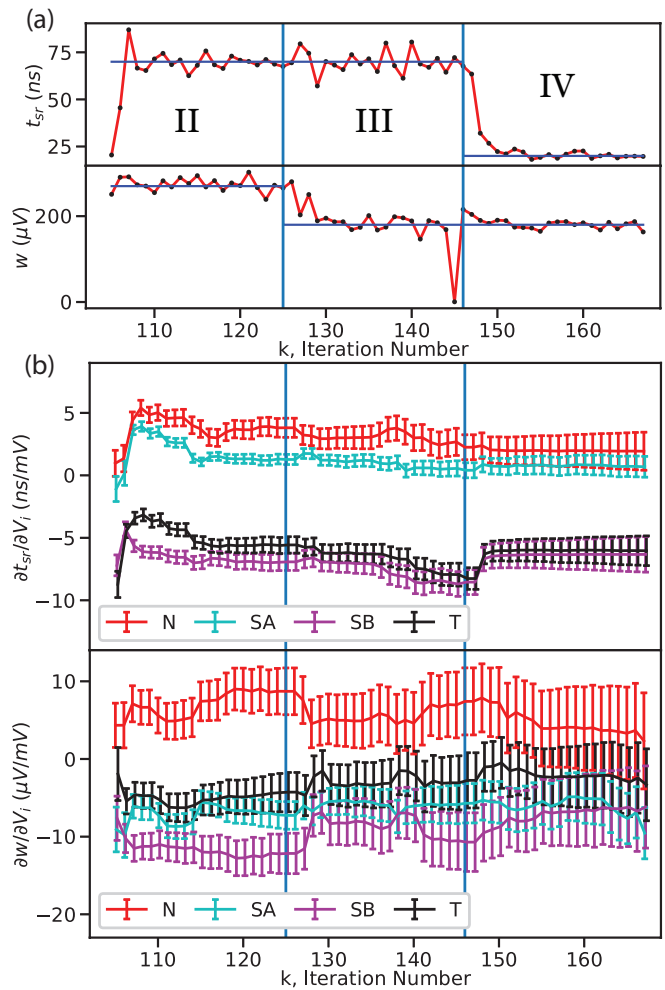}
    \caption{Continuation of the data shown in Fig.~\ref{fig::benchmarks}. The data is labeled in the same way as in Fig.~\ref{fig::benchmarks}. The complete demonstration took about 8 hours in lab time without any human interference. (a) The fast convergence of the parameters continues in the extension. At iteration number 145, the evaluation of $w$ fails because the hyperbolic tangent cannot be fitted to a noisy data set. (b) The gradient elements belonging to $w$ do not react to the failed fit at iteration number 145. This demonstrates the ability to quantify the measurement noise $\delta_w$ as discussed in Sec. \ref{sec::choice_of_q} and the ability of the Kalman filter to include the measurement noise in the update on the gradient elements. When Fig.~\ref{fig::benchmarks} is concatenated with the data shown in this figure, one can see that the variances sometimes shrink when the set point of the corresponding parameter value is changed. Otherwise the variance grows steadily as voltages are changed but little information is gained.}
    \label{fig::second_half}
\end{figure}

\section{Sensing Dot and Chemical Potentials}
\label{sec::sd_cds}
The measurement of $t_{\text{sr}}$ and $w$ requires a certain contrast in the spin-to-charge conversion and the exact knowledge of the chemical potentials.
Any time gate voltages on the DC-gates defining the double quantum dot are changed, we sweep either one or both of the voltages on the gates SD1 and SD2. The resulting CSD features Coulomb oscillations and we set the voltages of SD1 and SD3 to the steepest point on either side of the highest Coulomb peak to maximize the readout contrast.

The measurement of $t_{\text{sr}}$ and $w$ is performed by pulsing the RF-gates to different voltage points in the CSD of RFA and RFB\cite{Botzem2018}. The positions of these is chosen relative to the charge transition lines and can be kept fixed by compensating any shift of the transition lines caused by changes in the chemical potentials.


In the step "Set New Voltages" of the tuning process as described in Fig.~\ref{fig::KalmanSolver}, we use virtual gates - as discussed by Botzem \textit{et al.}\cite{Botzem2018} - to calculate a linear compensation on PA and PB to correct for the change in the chemical potential, caused by the change of the voltages on N, SA, SB and T. Subsequently, we correct for the non-linearity and inaccuracies in the compensation using an algorithm running in two loops. The inner loop tunes the sensing dot while the outer loop compensates shifts of the transition lines in the CSD. The outer loop employs the same Kalman-filter based tuning algorithm described in the main text by using the position of the charge transition lines in the CSD as the target parameters instead of $t_{\text{sr}}$ and $w$.


Tab.~\ref{tab:time_consumption} lists the time consumption of each measurement mentioned above. For a detailed discussion of the tuning performance and possible improvements, we give also the number of times each measurement has been conducted during the experimental demonstration and the resulting fractional share of the total measurement time. Thereby, we identify the maximization of the contrast in the sensing dot and the compensation of shifts in the chemical potential as bottleneck of the tuning procedure. They require the largest share of the measurement time because they are performed much more often than the extraction of the parameters $t_{\text{sr}}$ and $w$. The measurements were not optimized for short measurement times.



\begin{table}[h]
    \centering
    \begin{tabular}{c|r|r|r}
         Measurement & $t_{\text{Extraction}}$(\SI{}{\second}) & $N_{\text{Evaluations}}$ & $t_{\text{Fraction}}$(\%) \\ \hline
         SD 1DIM & 17 & 532 & 31 \\
         SD 2DIM & 100 & 5 & 2 \\
         Chem. Pot. & 38 & 335 & 49 \\
         $w$ & 28 & 168 & 16 \\
         $t_{\text{sr}}$ & 5 & 168 & 3 \\
    \end{tabular}
    \caption{Time consumption of the parameter extraction. The one and two dimensional sweeps of the sensing dot gates SD1 and SD2 and the measurement of the chemical potentials are abbreviated by SD 1DIM, SD 2DIM and Chem. Pot. respectively.  $t_{\text{Extraction}}$ is the time required for a single measurement and $N_{\text{Evaluations}}$ is the total number of times a measurement has been executed during the experimental demonstration. The fractional share of the total time elapsed during the experimental demonstration is given by $t_{\text{Fraction}}$.}
    \label{tab:time_consumption}
\end{table}

\section{Initial conditions}
\label{sec::init}
In the initial step ($k=0$),
the gradient is measured several times by finite differences and $\boldsymbol{g}^{(0)}$ and $\boldsymbol{C}^{(0)}$ are calculated by averaging over these measurements. $\boldsymbol{C}^{(0)}$ is a diagonal matrix having as elements the variances of the measured gradients. The duration of this measurements depends on the number of measurements at each point in voltage space. Within 60 min, decent approximations can be measured. Once calculated, the gradient can be used for all subsequent tuning routines.

Even if only a rough estimate of $\boldsymbol{g}^{(0)}$ is available, the algorithm can be adjusted by choosing $\boldsymbol{C}^{(0)}$ in the same order of magnitude as the squares of the elements in $\boldsymbol{g}^{(0)}$. In this case the gradient will be effectively remeasured in the tuning process while profiting from the guessed information in $\boldsymbol{g}^{(0)}$.

\section{Measurement Uncertainty}
\label{sec::measurement_noise}
The largest issue to the stability of the tuning procedure during the experimental demonstration arises from statistical noise on the raw measurement data. 
To evaluate $t_{sr}$ and $w$ we apply a pulse while sweeping one pulse parameter $x$ and measuring the signal $y$ via RF-reflectometry. This yields a raw data set consisting of measurement points $(x_i,y_i)$ with $ 1 \leq i  \leq n$. We fit a function $f_p(x)$ to extract the parameter $p$ being $t_{sr}$ or $w$\cite{Botzem2018}.

We quantify the statistical noise on a parameter fit with the following procedure. 
After fitting $f_p$ to the data, we calculate the quadratic sum of the residuals
\begin{equation}
    \sigma = \sum_i \left( \frac{f_p(x_i)-y_i}{\Theta_p}\right)^2.
\end{equation}
where $\Theta_p$ is a rescaling factor describing the sensitivity in the sensing dot. $\Theta_w$ is chosen as height of the inter-dot transition and $\Theta_{t_{\text{sr}}}$ is chosen as range of $(y_1, \dots, y_n)$, i.e. both describe the response in the sensing dot signal to the transition of an electron within the double quantum dot. 

$\sigma$ is averaged over many measurements before the tuning to calculate a reference value $\sigma_0$. The standard deviation of this set of measurements is denoted by $\delta p_0$ and we estimate the uncertainty due to statistical errors as $\delta p \approx \delta p_0 \sigma / \sigma_0$. 

\section{Choice of $\boldsymbol{Q}$}
\label{sec::choice_of_q}
A good choice of the parameter uncertainty added in each step,  $\boldsymbol{Q}$, is crucial for the performance and stability of the Kalman filter as it controls the Kalman gain.
We used for the singlet reload time $\boldsymbol{Q}_{t_{\text{sr}}} = (\SI{0.075}{\nano \second \per \milli \volt})^2\cdot \boldsymbol{I}$ and for the transition width $\boldsymbol{Q}_{w} = (\SI{0.3}{\micro \volt \per \milli \volt})^2\cdot \boldsymbol{I}$. 
For an update step of maximal step size (\SI{10}{\milli \volt}) $\boldsymbol{H}\boldsymbol{Q} \boldsymbol{H}^T = \text{diag}((\SI{0.75}{\nano \second})^2, (\SI{3}{\micro \volt})^2)$ is of the same order of magnitude as typical values for the measurement noise $(\delta t_{\text{sr}}^2, \delta w^2) $ as estimated from repeated measurements at fixed gate voltages. However, in most update steps the voltage change is less than \SI{10}{\milli \volt} yielding lower values in $\boldsymbol{H}$ so that the presence of $\Delta z^2$ in the denominator of Eq. \eqref{eq::kalman_gain} leads to a substantially smaller adjustment of the gradient estimated than would be obtained for perfect measurements with $\Delta z^2=0$.

If $\boldsymbol{Q}$ is chosen too small, the procedure leads to a slow adaptation of the gradient based on new measurements in each iteration, which may lead to the same disadvantages as using Eq. \eqref{eq::gaussnewton} with constant gradients. These include slow convergence to the target values if the gradient elements are estimated to large, and oscillation or divergence if the gradient elements are estimated to small so that each step overcompensates the deviation from the target values.
On the other hand, a large $\boldsymbol{Q}$ can be problematic if individual measurements sometimes produce outliers, e.g., due to undetected bad fit convergence, as these then lead to a large erroneous change in the gradient that may compromise the continuation of the algorithm. Furthermore, previously available information is then discarded very quickly. Smaller $\boldsymbol{Q}$ dampen the response to such events and make better use of the information in the history of previous iterations. Hence, one must choose $Q$ just large enough so that excessive oscillations are avoided. 
A potential improvement would be to choose $\boldsymbol{Q}^{(k)}$ such that it increases with $v^{(k)}-v^{(k-1)}$ so that larger adjustments are made to the gradient if larger changes are to be expected whereas information is retained if can be expected to be valuable.

\end{document}